\documentclass[12pt]{article}
\usepackage[dvips]{graphicx}
\usepackage{latexsym}

\newcommand{\scalar}	{{\rm S}}
\newcommand{\elemag}	{{\rm EM}}
\newcommand{\ELE}	{{\rm E}}
\newcommand{\MAG}	{{\rm M}}
\newcommand{\DOF}	{{\rm dof}}

\newcommand{\Hamil}	{{\cal H}}
\newcommand{\fs}	{{\rm 1s}}

\newcommand{\sing}	{{\rm sing}}

\newcommand{\BH}	{{\rm BH}}
\newcommand{\pl}	{{\rm pl}}

\newcommand{\fig}[1]	{Figure \ref{#1}}
\newcommand{\EQ}[1]	{(\ref{#1})}

\begin{document}

\begin{flushright}
 \begin{minipage}[b]{43mm}
  hep-th/0501021\\
  OIQP-04-07\\
 \end{minipage}
\end{flushright}

\renewcommand{\thefootnote}{\fnsymbol{footnote}}
\begin{center}
 {\Large\bf
 A Structure Model for Black Holes:\\
 Atomic-like Structure, Quantization \\
 and the Minimum Schwarzschild Radius
 }\\
 \vspace*{3em}
 {Yukinori Nagatani}\footnote
 {e-mail: yukinori\_nagatani@pref.okayama.jp}\\[1.5em]
 {\it Okayama Institute for Quantum Physics,\\
 1-9-1 Kyoyama, Okayama City 700-0015, JAPAN}
\end{center}
\vspace*{1em}

\begin{abstract}
 A structure model for black holes is proposed
 by mean field approximation of gravity.
 The model,
 which consists of a charged singularity at the center and
 quantum fluctuation of fields around the singularity,
 is similar to the atomic structure.
 The model naturally quantizes the black hole.
 Especially
 we find the minimum black hole,
 whose structure is similar to the hydrogen atom 
 and
 whose Schwarzschild radius becomes
 about $1.1287$ of the Planck length.
\end{abstract}

\newpage
\section{Introduction}\label{intro.sec}

The most natural solution for the information paradox of black holes
is constructing a structure model of the black holes.
When we find a spherical object
whose properties,
e.g., mass, charges, radius, entropy \cite{Bekenstein:1973ur},
radiation from it \cite{Hawking:1975sw,Hawking:1974rv} and so on,
correspond with these of a black hole,
the object can be regarded as the black hole
for a observer who is distant from the black hole.
In other words,
the object becomes a model of the black hole.
The D-brane description of the (near) extremal-charged black holes
\cite{Strominger:1996sh,Horowitz:1996fn}
is one of the successful models.
If the object has no horizon
and has an interior structure instead of the inside of the horizon,
there arises no information paradox
\cite{Hotta:1997yj,Iizuka:2003ad,Nagatani:2003rj,Nagatani:2003new,Nagatani:2003new2}.
The object is just the structure model of the black holes.

K.~Hotta proposed the Planck solid ball model \cite{Hotta:1997yj},
which is just the structure model of black holes.
According to the Planck solid ball model,
any Schwarzschild black hole
consists of a ball of the Planck solid and
a layer of thermal radiation-fluid around the ball.
The Planck solid is a hypothetical matter
which arises by the stringy thermal phase transition
due to the high temperature of the radiation.
The temperature of the radiation around the ball
becomes very high
because
a deep gravitational potential (small $g_{00}(r)$)
makes the blue-shift effect ($T(r) \propto 1/\sqrt{g_{00}(r)}$)
\cite{Hotta:1997yj,Iizuka:2003ad,Nagatani:2003ps}.
Both the entropy and the mass of the black hole
are carried by the radiation.
The Hawking radiation is explained as a leak of the radiation
from the radiation-layer of the ball.

Based on the Planck solid ball model,
the radiation ball model for black holes is proposed
by considering the gravitational backreaction into the radiation
without the assumption of the Planck-solid-transition
\cite{Nagatani:2003rj,Nagatani:2003new,Nagatani:2003new2}.
In the radiation ball model,
any black hole consists of a central singularity and
a ball of radiation-fluid around the singularity.
The radius of the ball corresponds with the Schwarzschild radius.
The model is defined on the Einstein's static universe $(\Lambda > 0)$
of the radiation-dominance
because there is thermal equilibrium
between the ball and the background universe.
A lough estimation of entropy of the radiation ball
is proportional to the surface area of the ball and
is almost the same as the Bekenstein entropy.
The model also succeeded in explaining several properties of the
charged black holes including (near) extremal black holes.

To clarify the concept and redefine the structure models,
we propose a decomposition of fields as following.
In the theory which we are considering
there are
a gravity field (metric) $g_{\mu\nu}$,
a electromagnetic field $A_\mu$,
a scalar field $\phi$ and so on.
We decompose any field in the theory 
into the spherically-symmetric time-independent part
and the fluctuating part:
\begin{eqnarray}
 &&
 \begin{array}[c]{ccccc}
         g_{\mu\nu}(t,r,\theta,\varphi) &=& g_{\mu\nu}(r) &+&
  \delta g_{\mu\nu}(t,r,\theta,\varphi) \\
         A_{\mu}(t,r,\theta,\varphi)    &=& A_{\mu}(r)    &+&
  \delta A_{\mu}(t,r,\theta,\varphi)    \\
         \phi(t,r,\theta,\varphi)       &=& 0             &+&
  \delta \phi(t,r,\theta,\varphi)       \\
  \vdots                                & & \vdots        & &
  \vdots                                \\
  & & \mbox{\small Mean Field} & & \mbox{\small Fluctuation}
 \end{array},
 \label{decomposition.eq}
\end{eqnarray}
where $r,\theta$ and $\phi$ are the polar coordinates.
The time-independent parts are just mean fields.
The metric $g_{\mu\nu}(r)$ describes a background geometry
as a mean field.
Any field-fluctuation exists in the background space-time.
The electromagnetic field $A_{\mu}(r)$ also describes background
field.
The metric fluctuation $\delta g_{\mu\nu}$ describes gravitons
in the background.
The other fluctuations $\delta A_\mu, \delta \phi, \cdots$
correspond with the photons, the scalar bosons and so on.

Both in the Planck solid ball model and in the radiation ball model,
the temperature ansatz 
\begin{eqnarray}
 T(r) &\propto& 1/\sqrt{g_{00}(r)}
  \label{proper-temp.eq}
\end{eqnarray}
for the thermal radiation-fluid in the curved space-time $g_{\mu\nu}$
plays quite important role.
In these models the space-time structure
is described by the mean field of the gravity.
The fluid of the thermal radiation
is made of the particles as the field-fluctuations.
%
The temperature-ansatz \EQ{proper-temp.eq} is expected to be 
derived by a combination of
the mean field approximation and 
the fluid approximation of the particles.

Here it is natural that
we directly quantize the field-fluctuations
instead of considering
the fluid-approximation of 
the particles with the ansatz \EQ{proper-temp.eq}.
The aim of this paper is constructing a structure model for black holes
by using both the mean-field gravity
and the quantization of the fluctuating-fields.

Our proposal of the structure model for a black hole is the following:
We assume the Einstein gravity, the electromagnetic field
and several fields.
The black hole consists of both a singularity at the center
and fluctuating fields around the singularity.
The singularity has electric (magnetic) charge and
makes a spherically-symmetric time-independent electric (magnetic) field
around the singularity.
The charge of the singularity
makes the singularity gravitationally-repulsive
due to the Einstein equation.
The fluctuating fields contain all species of the fields in the theory.
The repulsive singularity, static electromagnetic field and
the fluctuating fields make 
a gravitational field as a mean field
which obeys the Einstein equation.
The gravitational field
binds the field-fluctuation into a ball.
The radius of the ball just indicates the Schwarzschild radius
of the corresponding black hole.
The exterior of the ball corresponds
with the ordinary charged black hole.
The shape of the field-fluctuation in the ball
is determined by
a quantization of the fluctuating fields
in the mean field of the gravity background.
The righthand side of the Einstein equation
is the expectation value
of the energy-momentum tensor for the quantum-fluctuating fields.

Our picture of the black hole is quite similar to
that of the atom
which consists of a charged nucleus and a quantized electron field
of several quanta.
The radius of the ball, namely, the Schwarzschild radius
is quantized
due to the quantization of the fluctuating field in the ball.
Especially we construct the minimum structure of the black hole.
The field-fluctuation in the minimal model
contains only one quantum of the 1s-wave mode.
The minimum model is quite similar to the hydrogen atom
which has a single electron of the 1s-wave function.
The radius of the minimum model, namely,
the minimum Schwarzschild radius becomes about $1.1287$ of the Planck length.
There is an analogy between the the minimum Schwarzschild radius
in our model and the Bohr radius of the hydrogen atom.

\section{Model Construction}\label{ackreaction.sec}

Interactions among the field-fluctuations are not important
to construct our model
because 
the essence of our model is a quantization of the field-fluctuations
in the gravity as a mean field.
As an approximation,
we switch off the interaction among the fields
except for the background gravity and
we adopt the classical background gravity as a mean field.
Adopting both
the mean field approximation and a self-consistent analysis,
we can perform a canonical quantization of the fluctuating-fields
and
we obtain the configuration of
both the gravity and the fluctuating quantum field.

\underline{Effective Action}

We adopt $g_\DOF$ kinds of real scalar fields
$\{\:\phi_i \:|\: i=1\sim g_\DOF\}$ and
we assume the scalar fields represent all of the field fluctuations
$\{\delta g_{\mu\nu},\: \delta A_{\mu},\: \delta\phi,\: \cdots \}$
in the theory as an approximation.
Our system is described by an effective model for 
a gravity (metric) $g_{\mu\nu}$,
an electromagnetic field $F_{\mu\nu}$
and the real scalar fields ${\phi_i}$.
Both the metric and the electromagnetic field are
time-independent spherically-symmetric mean field.
We assume the Einstein gravity, therefore,
the effective action for the model is naturally given by
\begin{eqnarray}
 S &=&
  \frac{1}{2 \kappa}
  \int d^4x \sqrt{-g} R
  \;+\; 
  \int d^4x \sqrt{-g}
  g^{\mu\alpha} g^{\nu\beta} \frac{1}{4} F_{\mu\nu} F_{\alpha\beta}
  \nonumber\\&&
  \;+\;
  \sum_i
  \int d^4x \sqrt{-g} g^{\mu\nu}
  \frac{1}{2} (\partial_\mu\phi_i) (\partial_\nu\phi_i)
\end{eqnarray}
with the spherically-symmetric time-independent metric
\begin{eqnarray}
 ds^2 &=&
  F(r) dt^2 \;-\; G(r) dr^2
  \;-\; r^2 d\theta^2 \;-\; r^2\sin^2\theta d\varphi^2.
  \label{metric.eq}
\end{eqnarray}
In the metric background \EQ{metric.eq},
the solution of the spherically symmetric static electromagnetic field
becomes
\begin{eqnarray}
  F_{\mu\nu} dx^\mu \wedge dx^\nu
  &=& E(r) \sqrt{FG} dt \wedge dr
  \;+\; B(r) r^2 \sin\theta d\theta \wedge d\varphi,\nonumber
\end{eqnarray}
where 
\begin{eqnarray}
  E(r) \;:=\; \frac{Q_\ELE}{r^2}, \qquad
  B(r) \;:=\; \frac{Q_\MAG}{r^2}
  \label{EM-Field.eq}
\end{eqnarray}
are the radial elements of the electric
and of the magnetic field respectively.
The constant $Q_\ELE$ and $Q_\MAG$ are the electric and the magnetic
charges at the central singularity $(r=0)$.
The non-trivial elements of the energy-momentum tensor for the
electromagnetic field \EQ{EM-Field.eq} are
\begin{eqnarray}
  T_{\elemag 0}^{\ \ \ \ 0} \ =\ 
  T_{\elemag r}^{\ \ \ \ r} \ =\ 
 -T_{\elemag \theta}^{\ \ \ \ \theta} \ =\ 
 -T_{\elemag \varphi}^{\ \ \ \ \varphi} \ =\ 
 \frac{Q^2}{8\pi}\frac{1}{r^4},
\end{eqnarray}
where we have defined $Q^2 := Q_\ELE^2 + Q_\MAG^2$.

We will quantize the scalar fields $\{\phi_i\}$ by an operator formalism
in the background of the gravity field \EQ{metric.eq}
and will construct a Fock space.
Let $\left|\Phi\right>$ be a state of the scalar fields
as an element of the Fock space.
The state $\left|\Phi\right>$ will describe the structure
of the scalar fields in our solution for the model of the black hole.
We require the state $\left|\Phi\right>$ should be
an eigenstate of the Hamiltonian for the scalar fields
because we want to construct a time-independent solution.
The eigenstate of the Hamiltonian keeps
the expectation value of the energy-momentum tensor time-independent.
Non-eigenstate of the Hamiltonian, e.g. a coherent state,
may be applicable to the analysis of the semi-resonance
of the black holes
by more complicated assumption of the metric.
The Einstein equation becomes
\begin{eqnarray}
 R_{\mu\nu} - \frac{1}{2} g_{\mu\nu} R
  &=& \kappa
  \left\{
   T_{\elemag\mu\nu}
   +
   \sum_i
   \left<\Phi| T_{\scalar_i\mu\nu} |\Phi\right>
  \right\}.
  \label{ein0.eq}
\end{eqnarray}
The energy momentum tensor for the scalar field becomes
\begin{eqnarray}
 T_{\scalar_i\mu\nu}
  &=&
  \frac{1}{2}
  \left(
   \partial_\mu\phi^\dagger_i \partial_\nu\phi_i
   +
   \partial_\nu\phi^\dagger_i \partial_\mu\phi_i
  \right)
   - \frac{1}{2} g_{\mu\nu} g^{\alpha\beta}
   \partial_\alpha\phi^\dagger_i \partial_\beta\phi_i
\end{eqnarray}
as an operator relation.
The state $\left|\Phi\right>$ should be chosen
so that the expectation value of the energy momentum tensor
$\sum_i \left<\Phi| T_{\scalar_i\mu\nu} |\Phi\right>$ becomes diagonal
because of a consistency of the metric assumption \EQ{metric.eq}
and the righthand-side of the Einstein equation \EQ{ein0.eq}.

\underline{Quantization of the fields}

Let us consider the quantization of the scalar fields
by an operator formalism.
We keep our eyes on one of the scalar fields $\{\phi_i\}$.
The conjugate momentum of the scalar field $\phi(t,r,\theta,\varphi)$
is defined as
\begin{eqnarray}
 \Pi
  &:=& \frac{\delta S}{\delta (\partial_0{\phi})}
  \;=\; \sqrt{\frac{G}{F}} r^2 \sin\theta \; \partial_0{\phi}.
\end{eqnarray}
The Hamiltonian of the scalar field becomes
\begin{eqnarray}
 \Hamil &:=& \left(\int dr d\theta d\varphi \; \Pi \dot{\phi} \right) - L 
   \ =\ 
   \int dr d\theta d\varphi \; \sqrt{FG} r^2 \sin\theta \nonumber\\
 && \qquad \times \frac{1}{2}
   \left[
     \frac{1}{G r^4 \sin^2\theta}	\Pi^2
   + \frac{1}{G}			(\partial_r \phi)^2
   + \frac{1}{r^2}			(\partial_\theta \phi)^2
   + \frac{1}{r^2 \sin^2\theta}		(\partial_\varphi \phi)^2
   \right].
   \label{Hamiltonian.eq}
\end{eqnarray}
The Hamiltonian \EQ{Hamiltonian.eq} corresponds with the other description
\begin{eqnarray}
 \Hamil &=& \int dr d\theta d\varphi \sqrt{FG} r^2 \sin\theta \ 
  T_{\scalar 0}^{\ \ 0}.
\end{eqnarray}
We employ a canonical quantization of the scalar field by the
equal-time commutation relation
\begin{eqnarray}
 &&
 \left[
  \phi(t,r_1,\theta_1,\varphi_1),
  \Pi(t,r_2,\theta_2,\varphi_2)
 \right] \ =\  i
 \delta(r_1 - r_2) \delta(\theta_1 -\theta_2) \delta(\varphi_1 - \varphi_2),
 \nonumber\\
 &&
 \left[
  \phi(t,r_1,\theta_1,\varphi_1),
  \phi(t,r_2,\theta_2,\varphi_2)
 \right] \ =\  0, \quad
 \left[
  \Pi(t,r_1,\theta_1,\varphi_1),
  \Pi(t,r_2,\theta_2,\varphi_2)
 \right] \ =\  0.
\end{eqnarray}%

Classical mode functions of the scalar field $\phi$ are written as
\begin{eqnarray}
 &&
  e^{-i\omega_n t} f_{nl}(r) \: Y_l^{\ m}(\theta,\varphi),\qquad
  e^{+i\omega_n t} f_{nl}(r) \: Y_l^{\ m}(\theta,\varphi)
  \label{modes.eq}
\end{eqnarray}
which satisfy the equation of motion
\begin{eqnarray}
 \frac{1}{\sqrt{g}} \partial_\mu
  \left[ \sqrt{g} g^{\mu\nu} \partial_\nu \phi \right] &=& 0
  \label{scalar_eom.eq}
\end{eqnarray}
with a suitable boundary condition for $r$.
The index $n$ is the principal quantum number
which specify the mode's frequency $\omega_n$,
the real function $f_{nl}(r)$ describes radial part of the mode function
and
$Y_l^{\ m}$ is the real spherically harmonic function
which is normalized as
$\int d\theta \sin\theta d\varphi \;
 Y_{l_1}^{\ m_1} Y_{l_2}^{\ m_2}
 = \delta_{l_1l_2} \delta_{m_1m_2}$.
Because of the equation of motion \EQ{scalar_eom.eq},
the radial function $f_{nl}(r)$ should satisfy
\begin{eqnarray}
 \sqrt{\frac{F}{G}} \frac{1}{r^2} \partial_r
  \left[ \sqrt{\frac{F}{G}} r^2 \partial_r f_{nl} \right]
  \;+\; \omega_n^2 f_{nl}
  \;-\; l(l+1) \frac{F}{r^2} f_{nl} &=& 0.
  \label{scalar_eom2.eq}
\end{eqnarray}

\underline{Boundary condition}

According to our proposal,
a black hole consists of both a singularity at the center
and a ball of a quantum-fluctuating fields around the singularity.
The radius of the ball corresponds
with the Schwarzschild radius $r_\BH$ of the black hole.
The exterior region $(r \gg r_\BH)$ of the ball
corresponds with the ordinary black hole.
Therefore 
the metric for the exterior region $(r \gg r_\BH)$
should correspond with the Reissner-Nordstr\"om (RN) metric:
\begin{eqnarray}
 F_\BH(r) &=&
  \left(1 - \frac{r_\BH}{r}\right)
  \left(1 - q^2 \frac{r_\BH}{r}\right),
  \nonumber\\
 G_\BH(r) &=& 1/F_\BH(r),
  \label{RN.eq}
\end{eqnarray}
where the parameter $q := Q/(m_\pl r_\BH)$ indicates
the charge of the black hole.
The parameter $q$ has a value from $0$ to $1$.
The extremal-charged black hole has $q=1$.

The Hawking temperature of the black hole in \EQ{RN.eq} becomes
\begin{eqnarray}
 T_\BH &=& \frac{1}{4\pi} \frac{1}{r_\BH} (1 - q^2).
 \label{Hawking-temp.eq}
\end{eqnarray}
In the radiation ball model,
the Hawking radiation is explained as a leak of the thermal radiation
from the radiation ball.
In our structure model,
the Hawking radiation corresponds with
the field-fluctuation around the ball.
We choose the quite near extremal-charged black hole ($q \rightarrow 1$)
to concentrate on analyzing the precise structure in the black hole
rather than the process of the Hawking radiation.
The absence of the Hawking radiation
from the quite near extremal-charged black hole,
namely,
the absence of the field-fluctuation around the ball
simplifies the analysis.
Therefore we require boundary condition $f_{nl} = 0$ for $r \gg r_\BH$.

There is a charged singularity at the center in our proposal.
The quantum fluctuation of the fields around the singularity
is neutral.
Therefore the charge of the ball equals to the charge of the singularity.
Because
the energy momentum tensor near the singularity
is dominated by the static electromagnetic field,
the metric near the center $(r \rightarrow 0)$ becomes
Reissner-Nordstr\"om (RN) type:
\begin{eqnarray}
 F_\sing(r)
  &=&
  A^2
  q^2 \left(\frac{r_\BH}{r}\right)^2, \label{Fsing}\\
 G_\sing(r)
  &=&
  \frac{1}{q^2} \left(\frac{r}{r_\BH}\right)^2. \label{Gsing}
\end{eqnarray}
The constant $A$ is the red-shift factor of the singularity.
The factor is caused by the condensation
of the energy of the fluctuating-field in the ball.
The condensation makes a deep gravitational potential.
According to the singularity of the 
charged radiation-ball solution\cite{Nagatani:2003new},
we put the factor as
\begin{eqnarray}
  A^2 := \alpha^2 \times \frac{(1 - q^2)^2}{12960 \sqrt{2\pi} m_\pl^4 r_\BH^4},
\end{eqnarray}
where $\alpha$ is a constant.
The factor $\alpha$ becomes 1
when we assume the same singularity
as that of the charged radiation-ball solution \cite{Nagatani:2003new}.
The factor $\alpha$ means the red-shift of the singularity
for the observer in the infinite distance.
The factor $\alpha$ cannot be determined by itself, therefore,
the factor parameterizes the solution.
We refer to $\alpha$ as the red-shift parameter of the singularity.

The boundary condition for the scalar field $\phi$
at the center ($r\rightarrow0$) is determined by
consistency between the Hamiltonian \EQ{Hamiltonian.eq}
and the mode functions \EQ{modes.eq}.
The Hamiltonian \EQ{Hamiltonian.eq} should correspond
with that of a harmonic oscillator of the frequency $\omega$
when the mode function \EQ{modes.eq} is substituted
into the Hamiltonian \EQ{Hamiltonian.eq}.
By employing a partial integration
after multiplying the equation of motion \EQ{scalar_eom2.eq}
by $f_{n_2 l_2}$,
we find the following relation
\begin{eqnarray}
  && \int_0^\infty dr \sqrt{\textstyle \frac{F}{G}}
  (\partial_r f_{n_1l_1}) (\partial_r f_{n_2l_2})
  \ =\ 
  \left[
   \sqrt{\textstyle \frac{F}{G}} r^2 (\partial_r f_{n_1l_1}) f_{n_2l_2}
  \right]_0^\infty
  \nonumber\\
  && \qquad \;+\;
  \omega_{n_1}^2
  \int_0^\infty dr \sqrt{\textstyle \frac{G}{F}} r^2 f_{n_1l_1} f_{n_2l_2}
  \;-\;
   l_1(l_1+1)
   \int_0^\infty dr \sqrt{FG}  f_{n_1l_1} f_{n_2l_2}.
  \label{Integral-Relation.eq}
\end{eqnarray}
By using 
the relation \EQ{Integral-Relation.eq}
with $n_1 = n_2$ and $l_1 = l_2$ 
and by using
the relation in the spherical harmonic function
$\int d\theta \sin\theta d\varphi
\left[
 (\partial_\theta Y_l^{\ m})^2
 + \frac{1}{\sin^2\theta} (\partial_\varphi Y_l^{\ m})^2
\right] = l(l+1)$,
the Hamiltonian for the function
$\phi(t,r,\theta,\varphi) = A(t) f_{nl}(r) Y_l^{\ m}(\theta,\varphi)$
with arbitrary $n,l$ and $m$
becomes
\begin{eqnarray}
 \Hamil[\phi] &=& \frac{1}{2} C_{nl}
  \left[
   \dot{A}^2 +
	\left(
	 \omega_n^2 +
	 \frac{\left[
	  \textstyle \sqrt{\frac{F}{G}} r^2 (\partial_r f_{nl})f_{nl}
	 \right]_0^\infty}{C_{nl}}
	\right)
   A^2	
  \right],
  \label{Hamiltonian-A.eq}
\end{eqnarray}
where we have defined a constant as
\begin{eqnarray}
 C_{nl}
  &:=&
  \int_0^\infty dr \sqrt{\textstyle \frac{G}{F}} r^2 f_{nl}^2.
  \label{C.eq}
\end{eqnarray}
The constant \EQ{C.eq} indicates the square of the norm of $f_{nl}$.
For any $n$ and $l$,
$\left[
	\sqrt{\frac{F}{G}} r^2 (\partial_r f_{nl})f_{nl}
\right]_0^\infty
$
should banish
because the Hamiltonian \EQ{Hamiltonian.eq}
should describe a harmonic oscillator of frequency $\omega_n$.
Therefore
we find that the boundary condition at $r=0$ should be
$f_{nl} = 0$ or $\partial_r f_{nl} = 0$.

\underline{Creation and Annihilation Operators}

Because the relation \EQ{Integral-Relation.eq}
has a reparametrization symmetry $(n_1,l_1) \leftrightarrow (n_2,l_2)$,
we find
\begin{eqnarray}
  (\omega_{n_1}^2 - \omega_{n_2}^2)
  \int_0^\infty dr \sqrt{\textstyle \frac{G}{F}} r^2 f_{n_1l_1} f_{n_2l_2}
  &=&
   \left\{
    l_1(l_1+1) - l_2(l_2+1)
   \right\}
   \int_0^\infty dr \sqrt{FG}  f_{n_1l_1} f_{n_2l_2}
  \label{Integral-Relation2.eq}
\end{eqnarray}
and there arises an orthogonal relation for radial mode functions
$\{f_{nl}\}$ for the same $l$ as
\begin{eqnarray}
 \int_0^\infty dr \sqrt{\textstyle \frac{G}{F}} r^2 f_{n_1l_1} f_{n_2l_2}
  &=& 0 \qquad (l_1 = l_2 \mbox{\ and\ } \omega_{n_1} \neq \omega_{n_2})
\end{eqnarray}
from \EQ{Integral-Relation2.eq}.

We define an inner product for scalar fields on any time-slice as
\begin{eqnarray}
 (A, B) &:=&
  \int dr \sqrt{\textstyle \frac{G}{F}} r^2
  \int d\theta \sin\theta d\varphi
  \left[ A^* \partial_0 B - (\partial_0 A^*) B \right].
\end{eqnarray}
Among the mode functions \EQ{modes.eq},
there arises the following orthogonal relations:
\begin{eqnarray}
 \left(
  e^{-\omega_{n_1}} f_{n_1l_1} Y_{l_1}^{\ m_1},\;
  e^{-\omega_{n_2}} f_{n_2l_2} Y_{l_2}^{\ m_2}
 \right)
 &=&
 + 2 \omega_{n_1} C_{n_1l_1}
 \delta_{n_1n_2}\delta_{l_1l_2}\delta_{m_1m_2}, \nonumber\\
 \left(
  e^{+\omega_{n_1}} f_{n_1l_1} Y_{l_1}^{\ m_1},\;
  e^{+\omega_{n_2}} f_{n_2l_2} Y_{l_2}^{\ m_2}
 \right)
 &=&
 - 2 \omega_{n_1} C_{n_1l_1}
 \delta_{n_1n_2}\delta_{l_1l_2}\delta_{m_1m_2}, \nonumber\\
 \left(
  e^{\pm\omega_{n_1}} f_{n_1l_1} Y_{l_1}^{\ m_1},\;
  e^{\mp\omega_{n_2}} f_{n_2l_2} Y_{l_2}^{\ m_2}
 \right)
 &=& 0.
\end{eqnarray}

The mode expansion for the operator of the scalar field
is naturally given by
\begin{eqnarray}
 \phi &=&
  \sum_{nlm}
  \frac{1}{\sqrt{2 \omega_n C_{nl}}}\;
  \left\{
  a_{nlm} \; e^{-i\omega_n t} 
  \;+\;
  a_{nlm}^\dagger \; e^{+i\omega_n t} 
  \right\}
  f_{nl}(r) \: Y_l^{\ m}(\theta,\varphi),
\end{eqnarray}
where the summation is performed over all of the physical modes $\{nlm\}$
in the background of the gravity \EQ{metric.eq}.
The annihilation operator $a_{nlm}$
and the creation operator $a_{nlm}^\dagger$
are written as
\begin{eqnarray}
 a_{nlm}
  &=&
  \left(
   \textstyle \frac{1}{\sqrt{2 w_n C_{nl}}}
   e^{-i\omega_n t} f_{nl} Y_l^{\ m}, \phi
  \right) \nonumber\\
 &=&
  \frac{e^{+i\omega_n t}}{\sqrt{2 w_n C_{nl}}}
  \int dr \sqrt{\frac{G}{F}} r^2
  \int d\theta \sin\theta d\varphi f_{nl} Y_l^{\ m}
  \left[
   \omega_n \phi + i \sqrt{\frac{F}{G}}\frac{\Pi}{r^2 \sin\theta}
  \right],
  \nonumber\\
 a_{nlm}^\dagger
 &=&
  - \left(
   \textstyle \frac{1}{\sqrt{2 w_n C_{nl}}}
   e^{+i\omega_n t} f_{nl} Y_l^{\ m}, \phi
  \right) \nonumber\\
 &=&
  \frac{e^{-i\omega_n t}}{\sqrt{2 w_n C_{nl}}}
  \int dr \sqrt{\frac{G}{F}} r^2
  \int d\theta \sin\theta d\varphi f_{nl} Y_l^{\ m}
  \left[
   \omega_n \phi - i \sqrt{\frac{F}{G}}\frac{\Pi}{r^2 \sin\theta}
  \right].
\end{eqnarray}
Then we find the well-known commutation relations:
\begin{eqnarray}
 \left[a_{n_1l_1m_1}, a_{n_2l_2m_2}^\dagger\right]
  &=& \delta_{n_1n_2} \delta_{l_1l_2} \delta_{m_1m_2},
  \nonumber\\
 \left[a_{n_1l_1m_1}, a_{n_2l_2m_2} \right] &=& 0, \qquad
 \left[a_{n_1l_1m_1}^\dagger, a_{n_2l_2m_2}^\dagger \right] \ =\ 0.
\end{eqnarray}
The Hamiltonian \EQ{Hamiltonian.eq} becomes
\begin{eqnarray}
 \Hamil
 &=&
  \sum_{nlm} \frac{\omega_n}{2}
  \left\{
   a_{nlm} a_{nlm}^\dagger
   + 
   a_{nlm}^\dagger a_{nlm}
  \right\}.
  \label{Hamiltonian-AA.eq}
\end{eqnarray}
The vacuum of the system is defined as $a_{nlm} \left|0\right> = 0$
for all physical modes $\{nlm\}$.
The Fock space is constructed
by acting the creation operators $\{a^\dagger_{nlm}\}$
into the vacuum $\left|0\right>$.

Here we restore the index of the scalar fields $\{\phi_i\}$.
Our structure model for the black hole is constructed by
specifying the state $\left|\Phi\right>$
which is an eigenstate of the Hamiltonian \EQ{Hamiltonian-AA.eq}
and which keeps the expectation value of the energy momentum tensor
$\sum_i \left<\Phi| T_{\scalar_i\mu\nu} |\Phi\right>$ 
to diagonal.
When we find a solution of $F(r), G(r)$ and $\{f_{inl}(r)\}$
and also find constants $r_\BH$ and $\{\omega_{i}\}$
which satisfy the boundary conditions,
the state $\left|\Phi\right>$ and space-time structure
is specified.
In the next section, we concretely construct a model.

\section{The Minimal Structure Model for Black Hole}\label{model.sec}

We will find the minimal structure of the black hole.
The simplest structure consists of
only one quantum in the 1s-wave function $\left|\fs\right>$
which is defined as
\begin{eqnarray}
 \left|\fs\right> &:=& a_{i,n=1,l=0,m=0} \left|0\right>
\end{eqnarray}
for only one flavor $i$ of the scalar fields.
We keep our eyes on the quantum $i$ and omit to express it again.
The structure of the minimal model will be similar
to that of the hydrogen atom which consists of a proton at the center
and a single electron of the 1s-wave function.
The entropy of the simplest structure becomes $\log g_\DOF$
because
there is $g_\DOF$ flavors of the scalar fields in the theory
and the single quantum belongs to one of the $g_\DOF$ flavors.

Here a problem of the vacuum-energy arises,
namely the vacuum-energy of the fluctuating-fields diverges
because the zero-point energy of each mode contributes
to the vacuum energy.
To avoid the problem,
we assume a supersymmetry of the vacuum $\left|0\right>$
or we ignore the zero-point energy:
\begin{eqnarray}
 \left<0\right| \Hamil \left|0\right> &=& 0, \qquad
 \left<\fs\right| \Hamil \left|\fs\right>
 \ = \
 \frac{\omega_1}{2}
 \left<\fs\right|
	a_{100}^\dagger a_{100} + a_{100} a_{100}^\dagger
 \left|\fs\right>
 \ = \
 \omega_1.
\end{eqnarray}
Another solution for the problem
is the regularization of the vacuum energy.
We expect an effect of the Casimir energy into the structure
if we adopt the regularization.
The effect is interesting and may be a future subject.

The expectation value of the energy momentum tensor for the state
$\left|\fs\right>$
becomes
\begin{eqnarray}
 \left<\fs\right| {\textstyle T_0^{\ 0}} \left|\fs\right>
  \ =\ 
 -\left<\fs\right| {\textstyle T_r^{\ r}} \left|\fs\right>
  &=&
  \frac{1}{8\pi} \frac{1}{\omega_1 C_{10}}
  \left[
   \frac{\omega_{1}^2}{F} f_{10}^2 \;+\; \frac{1}{G} (\partial_r f_{10})^2
  \right],
  \nonumber\\
 -\left<\fs\right| {\textstyle T_\theta^{\ \theta}} \left|\fs\right>
  \ =\ 
 -\left<\fs\right| {\textstyle T_\varphi^{\ \varphi}} \left|\fs\right>
  &=&
  \frac{1}{8\pi} \frac{1}{\omega_1 C_{10}}
  \left[
   \frac{\omega_{1}^2}{F} f_{10}^2 \;-\; \frac{1}{G} (\partial_r f_{10})^2
  \right],\nonumber\\
  \left<\fs\right| (\mbox{non-diagonal-parts}) \left|\fs\right> &=& 0.
   \label{1s-EM-tensor.eq}
\end{eqnarray}
The Einstein equation \EQ{ein0.eq} becomes the following three equations:
\begin{eqnarray}
 \frac{-G + G^2 + r G'}{r^2 G^2}
  &=&
  \frac{8\pi}{m_\pl^2}
  \left\{
   +
   \frac{Q^2}{8\pi}\frac{1}{r^4} \;+\;
   \left<\fs\right| T_0^{\ 0} \left| \fs \right>
  \right\},
   \label{ein1.eq}
   \\
 \frac{F - FG + r F'}{r^2 F G}
  &=&
  \frac{8\pi}{m_\pl^2}
  \left\{
   -
   \frac{Q^2}{8\pi}\frac{1}{r^4} \;-\;
   \left<\fs\right| T_r^{\ r} \left| \fs \right>
  \right\},
   \label{ein2.eq}
   \\
  \frac{- r G F'^2 - 2 F^2 G' - F \{r F' G' - 2 G (F' + r F'')\}}
   {4 r F^2 G^2}
  &=&
  \frac{8\pi}{m_\pl^2}
  \left\{
   +
   \frac{Q^2}{8\pi}\frac{1}{r^4}
   \;-\;
   \left<\fs\right| T_\theta^{\ \theta} \left| \fs \right>
  \right\},
  \label{ein3.eq}
\end{eqnarray}
where the prime means a differentiation by $r$.
By subtracting \EQ{ein2.eq} from \EQ{ein1.eq}
with a property
$ \left<\fs\right| T_0^{\ 0} \left| \fs \right>
= - \left<\fs\right| T_r^{\ r} \left| \fs \right>$,
we find a relation
\begin{eqnarray}
 G &=& \frac{\frac{r}{2} \frac{H'}{H} + 1}{1 - \frac{Q^2}{m_\pl^2 r^2}},
  \label{defG.eq}
\end{eqnarray}
where we have defined a new parameter
\begin{eqnarray}
 H &:=& \frac{F}{G}.
  \label{defH.eq}
\end{eqnarray}

By the relation \EQ{defG.eq} and the definition \EQ{defH.eq},
the Einstein equation \EQ{ein1.eq} becomes
an equation for $H$:
\begin{eqnarray}
 &&
 \frac{
  - 4\frac{Q^2}{m_\pl^2}\frac{1}{r^4} H^2
  -  \left(1 - 2\frac{Q^2}{m_\pl^2}\frac{1}{r^2}\right) H'^2
  +  \left(1 -  \frac{Q^2}{m_\pl^2}\frac{1}{r^2}\right)
     \left( \frac{4}{r} HH' + 2 HH''\right)
 }{(2H + rH')^2}
 \nonumber\\
 &=&
 \frac{8\pi}{m_\pl^2}
 \left\{
    \frac{Q^2}{8\pi}\frac{1}{r^4}
    \;+\;
    \frac{1}{8\pi}
    \frac{1 - \frac{Q^2}{m_\pl^2}\frac{1}{r^2}}{\frac{r}{2}H' + H}
    \;
    \frac{\omega_{1}^2 f_{10}^2 + H f_{10}'^{2}}{\omega_{1} C_{10}}    
 \right\}.
 \label{ein1-2.eq}
\end{eqnarray}

For convenience we define a normalized radial mode function as
\begin{eqnarray}
 \tilde{f} &:=& \frac{1}{\sqrt{\omega_{1}C_{10}}} f_{10},
\end{eqnarray}
which has a dimension of mass.
The normalization condition for $\tilde{f}$ is derived as
\begin{eqnarray}
  \omega_{1}
  \int_0^\infty dr \frac{1}{\sqrt{H}} r^2 \tilde{f}^2
  &=& 1
  \label{normalization.eq}
\end{eqnarray}
from the definition of $C_{10}$ in \EQ{C.eq}.
The Einstein equation \EQ{ein1-2.eq} becomes
\begin{eqnarray}
   \frac{2 H H'' - H'^2 + \frac{4}{r} H H'}{2 H + r H'}
   \;-\; 4 q^2 \frac{r_\BH^2}{r^4} \frac{H}{1 - q^2 \frac{r_\BH^2}{r^2}}
   \;-\; \frac{2}{m_\pl^2}
   \left( \omega_{1}^2 \tilde{f}^2 + H \tilde{f}'^2 \right)
   &=& 0.
\end{eqnarray}
The mode function $\tilde{f}$ should satisfy
the equation of motion \EQ{scalar_eom2.eq} with $l=0$ because of s-wave.
Then the equation for $\tilde{f}$ becomes
\begin{eqnarray}
 H \tilde{f}''
 \;+\; \left(\frac{2}{r} H 
 \:+\: \frac{1}{2} H' \right) \tilde{f}'
 \;+\; \omega_{1}^2 \tilde{f}
 &=& 0.
 \label{scalar_eom3.eq}
\end{eqnarray}

Here we will check a consistency in
all of the Einstein equations
\EQ{ein1.eq}, \EQ{ein2.eq}, \EQ{ein3.eq} and
the equation of motion for scalar mode-function \EQ{scalar_eom3.eq}.
We find a relation
\begin{eqnarray}
 \frac{1}{r} \partial_r
 \left[
  \frac{r^2}{2} \left(\omega_{1}^2 \tilde{f}^2 + H \tilde{f}'^2\right)
 \right]
 &=& \omega_{1}^2 \tilde{f}^2 - H \tilde{f}'^2
 \label{frel.eq}
\end{eqnarray}
by multiplying both sides of the equation of motion \EQ{scalar_eom3.eq}
by $\tilde{f}'$.
The rest of the Einstein equation \EQ{ein3.eq},
which is not used in our argument,
is derived by applying the relation \EQ{frel.eq} into \EQ{ein1-2.eq}.
Therefore all of the Einstein equations
are consistent with 
the quantized scalar-field of the s-wave.

We have already considered the boundary condition
for the scalar field near the singularity, namely,
$\tilde{f} = 0$ or $\partial_r\tilde{f} = 0$ for $r \rightarrow 0$.
Since $\omega_{1}$ should be the lowest frequency
in order to construct the simplest structure model,
we choose $\partial_r\tilde{f} = 0$ 
as the boundary condition for $r \rightarrow 0$.

The simultaneous equations \EQ{ein1-2.eq} and \EQ{scalar_eom3.eq}
are numerically solved with the initial condition
\begin{eqnarray}
 &&
 F(r) \ \rightarrow\  F_\sing(r),\quad
 G(r) \ \rightarrow\  G_\sing(r),\quad
 \tilde{f}(r) \ \rightarrow\  \tilde{f}_\sing,\quad
 \tilde{f}'(r) \ \rightarrow\  0
\end{eqnarray}
for $r\rightarrow0$.
We should choose the radius of the solution $r_\BH$,
the initial value of the mode function $\tilde{f}_\sing$
and the frequency of the mode $\omega_{1}$
to satisfy the exterior boundary condition
\begin{eqnarray}
 &&
 F(r) \ \rightarrow\  F_\BH(r),\quad
 G(r) \ \rightarrow\  G_\BH(r),\quad
 \tilde{f}(r) \ \rightarrow\  0
\end{eqnarray}
for $r\gg r_\BH$ and
the normalization condition for $\tilde{f}$ in \EQ{normalization.eq}.
These boundary conditions makes a quantization of $r_\BH$.
There should arise no zero-cross of $\tilde{f}$ for $0<r<r_\BH$
because the function $\tilde{f}$ describes the lowest s-wave mode
($\fs$-mode).

We obtained the solution for $q = 0.99999$ and $\alpha=1$.
The resultant radius of the solution becomes
\begin{eqnarray}
 r_\BH &=& 1.12871 \times l_\pl,
\end{eqnarray}
where $l_\pl := 1/m_\pl$ is the Planck length.
The frequency of the mode becomes
\begin{eqnarray}
 \omega_{1} = 0.0253909 \times \alpha \: (1-q^2) \: m_\pl.
 \label{omega-result.eq}
\end{eqnarray}
The red-shift factor $\alpha$ of the singularity and
the factor $(1-q^2)$ which is the shift of the Hawking temperature
\EQ{Hawking-temp.eq}
are directly reflected in the frequency $\omega_{1}$.
The form of the radial mode-function for the scalar field is
displayed in \fig{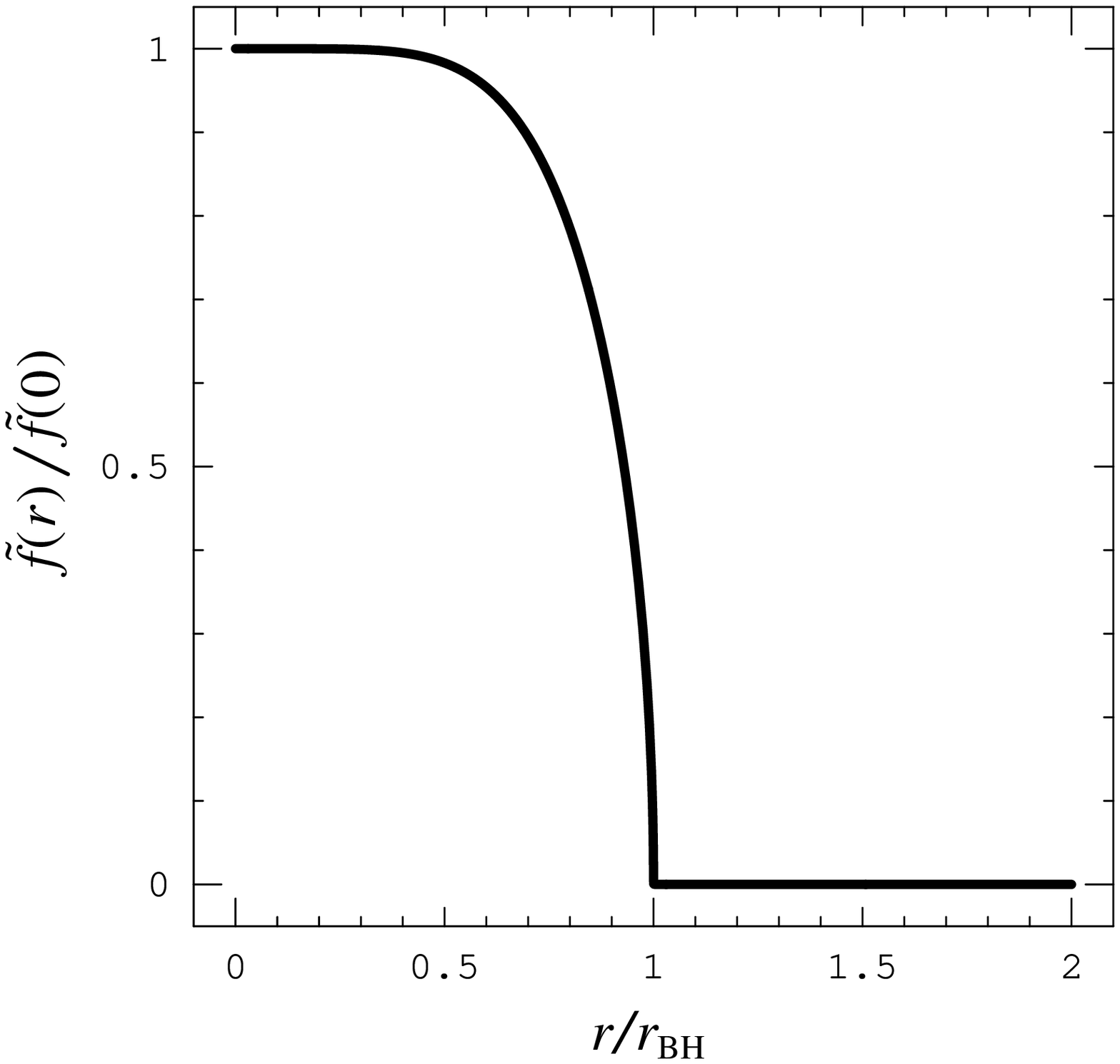},
where
the initial value of the radial mode-function is
\begin{eqnarray}
 \tilde{f}_\sing &=& \tilde{f}(0) \ =\  1.268776 \times m_\pl.
\end{eqnarray}%
The distributions of the resultant metric elements $F(r)$ and $G(r)$
are shown in \fig{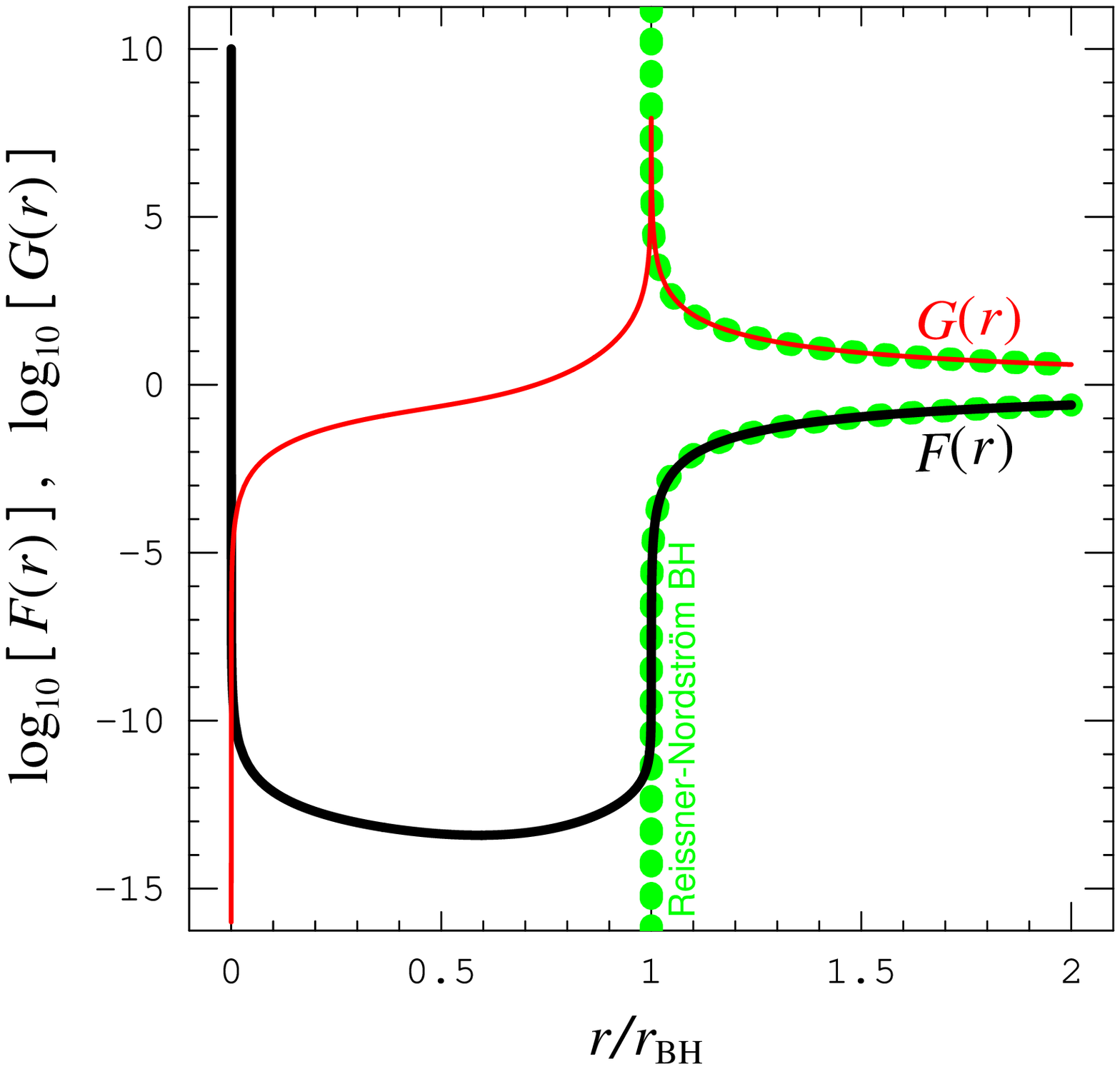}.
To contrast our result and the charged black hole,
the exterior part $(r>r_\BH)$
of the Reissner-Nordstr\"om metric \EQ{RN.eq}
of the same radius $r_\BH$ and the same charge $q$
are also indicated by the thick dotted (green) curves
in \fig{Fig2.FG.eps}.

\begin{figure}
 \begin{center}
  \includegraphics[scale=0.8]{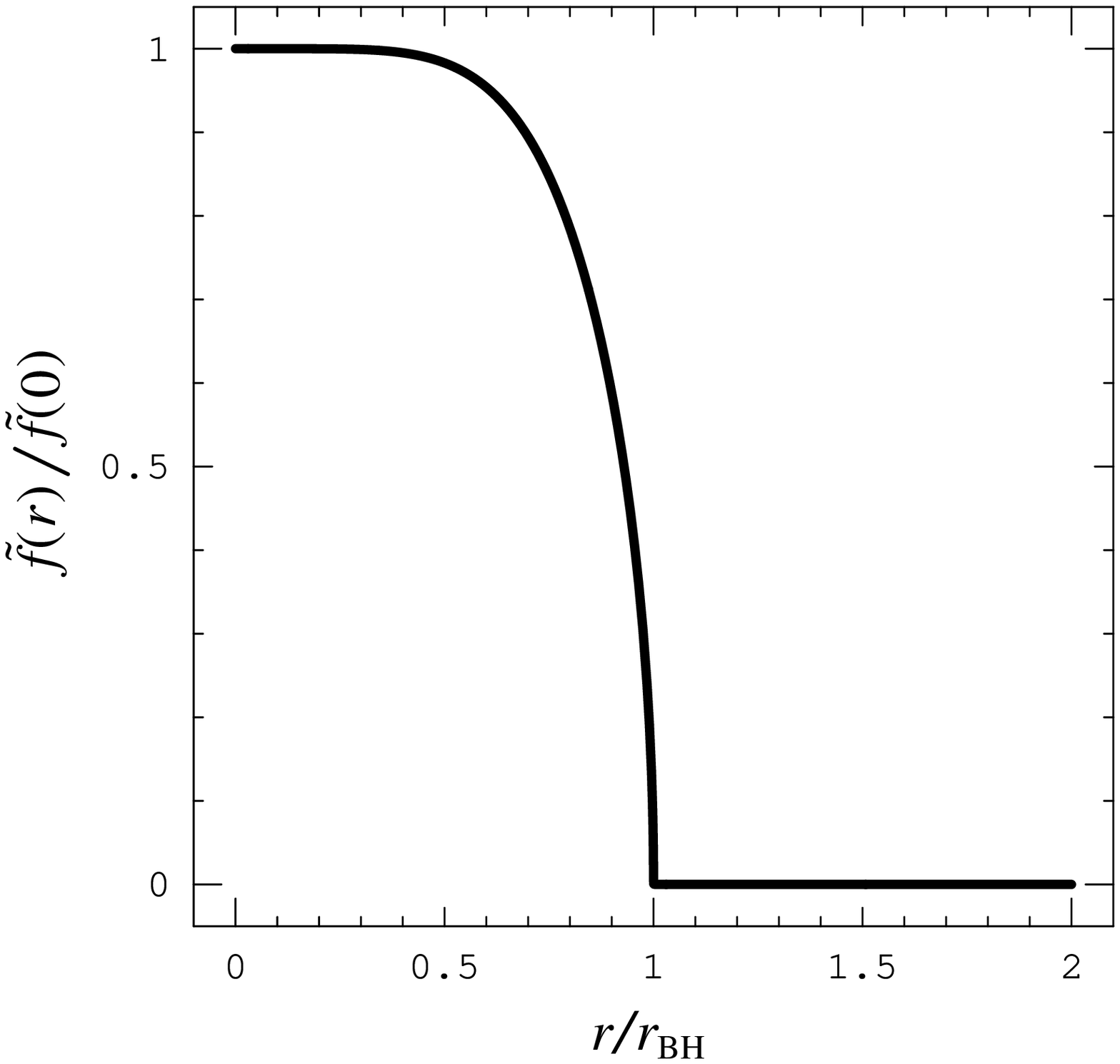}%
 \end{center}
 \caption{%
 Distributions of the radial mode-function $\tilde{f}(r)$
 for the $\left|\fs\right>$ scalar field in our solution
 with q=0.99999 and with $\alpha=1$.
 The vertical axis is normalized by $\tilde{f}(0) = 1.268776 \times m_\pl$.
 The horizontal axis is the coordinate $r$ normalized by $r_\BH$.
 The radius of the solution,
 namely the minimal Schwarzschild radius,
 becomes $r_\BH = 1.12871 \times l_\pl$,
 where $l_\pl := 1/m_\pl$ is the Planck length.
 \label{Fig1.f.eps}%
 }%
\end{figure}

\begin{figure}
 \begin{center}
  \includegraphics[scale=0.8]{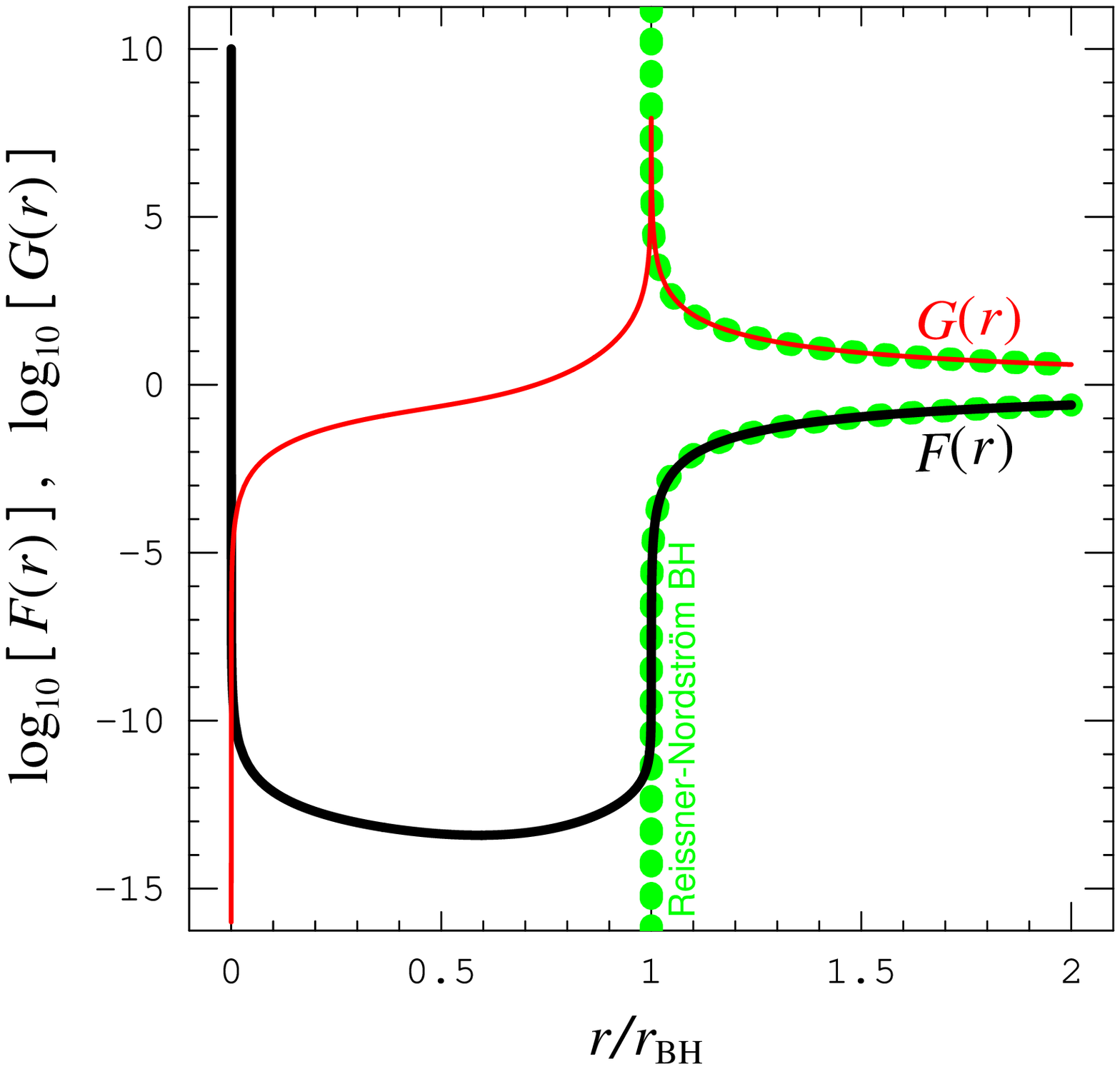}%
 \end{center}
 \caption{%
 Elements of the metric $F(r)$ and $G(r)$ in the solution.
 The thick solid (black) curve is $F(r)$ and
 the thin solid (red) curve is $G(r)$.
 The thick dotted (green) curves indicate
 the exterior part ($r>r_\BH$) of the Reissner-Nordstr\"om metric
 ($F_\BH(r)$ and $G_\BH(r)$ in \EQ{RN.eq})
 with the same radius $r_\BH$ and with the same charge $q$.
 \label{Fig2.FG.eps}%
 }%
\end{figure}

\begin{figure}
 \begin{center}
  \includegraphics[scale=0.8]{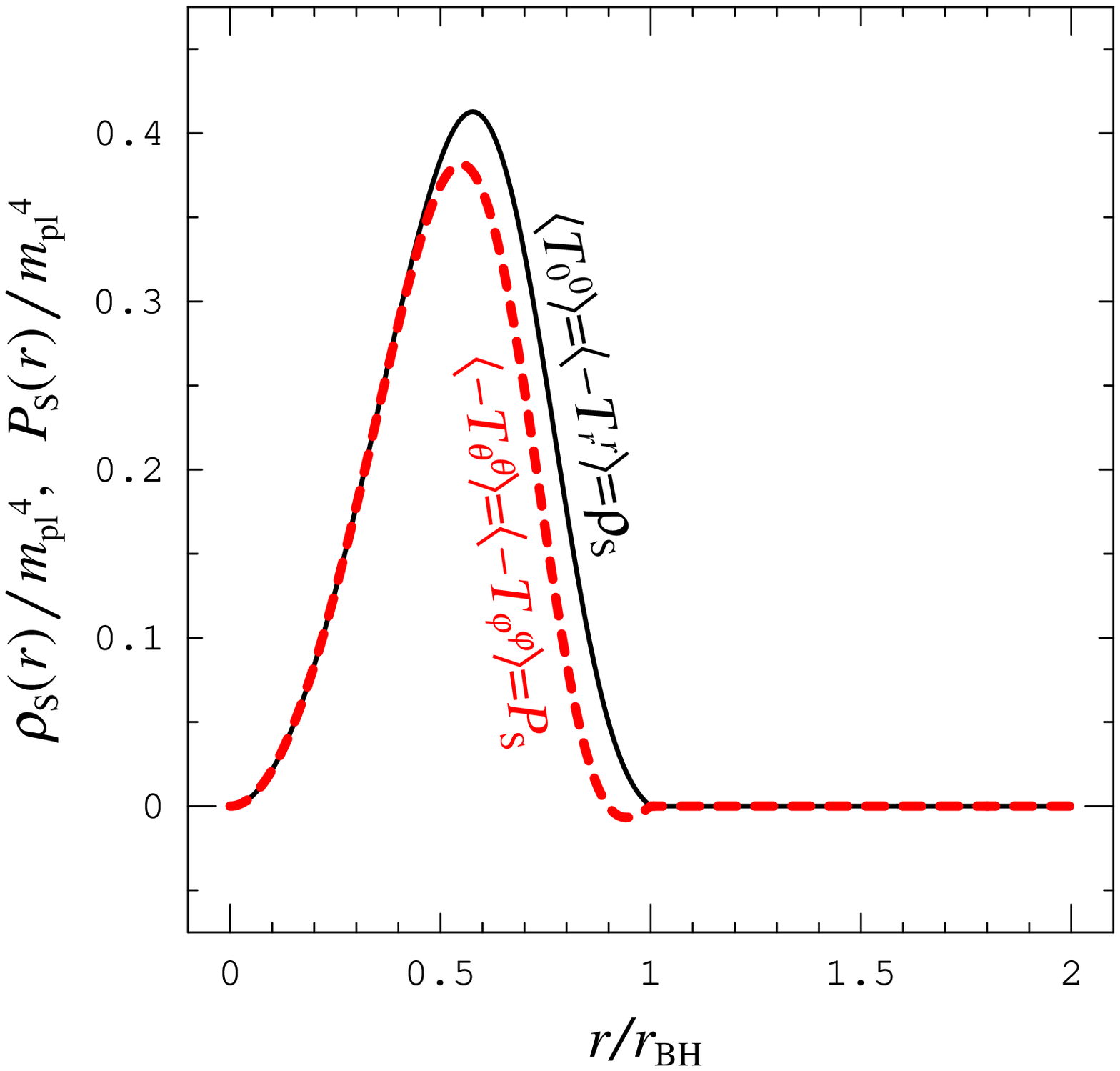}%
 \end{center}
 \caption{%
 Distribution of 
 the expectation values of the energy momentum tensor
 for the quantized scalar field
 $\left<\fs\left| T_{\scalar\mu}^{\ \ \nu} \right|\fs\right>$.
 The thin solid (black) curve is
 $\rho_\scalar
 =  \left<T_0^{\ 0}\right>
 = -\left<T_r^{\ r}\right>$.
 The thick dotted (red) curve is
 $P_\scalar
 =  \left<T_\theta^{\ \theta}\right>
 = -\left<T_\varphi^{\ \varphi}\right>$.
 $\rho_\scalar$ is positive definite but
 $P_\scalar$ becomes negative at $r \simeq r_\BH$.
 \label{Fig3.rho.eps}%
 }%
\end{figure}

According to the solution,
the quantum fluctuation of the scalar field is localized
in the ball whose radius corresponds with the Schwarzschild radius
(see \fig{Fig1.f.eps}).
Also the energy-density and the pressures of the scalar field
are localized in the ball (see \fig{Fig3.rho.eps}).
The shapes of the metric-elements in our solution
are qualitatively equivalent to
those in the radiation-ball solution
for charged black holes \cite{Nagatani:2003new}.
The form of the metric element $F(r)$
shows that there is a gravitational potential-well in the ball.
Therefore
the quantum fluctuation is binded into the ball
by the gravitation and
the gravitation is made by the singularity, the electromagnetic field
and the quantum fluctuation.
Everything in the exterior of the ball corresponds
with the charged black hole with the same charge $q$
except for the region extremely near the ball.
We conclude that the solution realizes
our proposal presented in the first section.

When we change the charge parameter
($q=0.99999, 0.9999, 0.999$)
within the range which can be regarded as near-extremal
or
when we change the red-shift parameter ($\alpha=1, 10, 100$),
there is almost no change in
the radius of the solution $r_\BH$,
the radial mode-function $\tilde{f}(r)$
and
the expectation value of the energy-momentum tensor
$\left<\fs\left| T_{\scalar\mu}^{\ \ \nu}(r) \right|\fs\right>$.
When $\alpha$ becomes much larger or when $q$ becomes much smaller,
the border of the ball becomes fuzzy.
When $q$ approaches to 1 (extremal charge) or
when $\alpha$ becomes small,
the potential-well ($F(r)$ in the ball) becomes deep
with keeping both $\tilde{f}(r)$ and
$\left<\fs\left| T_{\scalar\mu}^{\ \ \nu}(r) \right|\fs\right>$.
When the potential-well becomes deep,
the frequency $\omega_{1}$ becomes small
according to the dependency in \EQ{omega-result.eq}.

\section{Conclusion and Discussion}\label{discussion.sec}

We proposed a concept for the structure model of the black holes
by the mean field approximation.
The structure of the black hole,
which
consists of both a central singularity
and a ball of the quantum-fluctuation of the fields,
is similar to the atomic structure.
Especially
we concretely constructed the minimal structure model
for near extremal-charged black hole.
The minimal structure model contains a single quantum
of the field-fluctuation.
The structure of the minimal model is quite similar
to that of the hydrogen atom
and has the minimal Schwarzschild radius
$r_\BH = 1.12871 \times l_\pl$.

While our model describes the near extremal-charged black hole,
we expect that the model also describes the extremal limit
$(q \rightarrow 1)$.
The situation of our model is the same as
that of the radiation ball model for charged black hole
\cite{Nagatani:2003new}
where the extremal-charged black hole was expected to be described as
``the fully frozen radiation-ball''.
Our model of the extremal limit has
the same $\tilde{f}(r)$ in \fig{Fig1.f.eps} and
the same $\left<\fs\left| T_{\scalar\mu}^{\ \ \nu}(r) \right|\fs\right>$
in \fig{Fig3.rho.eps}.
The distribution of the metric element $G(r)$ in the extremal limit
corresponds with that in the non-extremal solution
except for the divergence of $G(r)$ on the peak
at $r=r_\BH$ (see \fig{Fig2.FG.eps}).
For the extremal limit,
the potential-well becomes infinitely deep
and the quantum-fluctuation freezes up
because the metric element $F(r)$ in the ball goes to zero
and the frequency $\omega_{1}$ also goes to zero.
The exterior region $(r>r_\BH)$ of our solution with the extremal limit
fully corresponds with that of the extremal-charged black hole.

The concrete construction of the non-minimal model
which contains multiple quanta is important subject.
This subject is technically complicated
because we should solve simultaneous equations
including unknown functions
of the same number as the quanta.
When we find the solution,
we can compare the entropy of the model and that
of the black hole.
The subject is related with analyzing
the process of the Hawking radiation in the model of
the black hole whose charge is much smaller than the extremal-charge.

The minimal model
cannot determine
the red-shift parameter $\alpha$ of the singularity by itself.
The radiation ball model
can determine the parameter $\alpha$ as $1$
because the model derives the Hawking radiation
as thermal equilibrium \cite{Nagatani:2003rj,Nagatani:2003new}.
Therefore analyzing the Hawking radiation from the ball
determines the parameter.

We adopted the approximation of representing
all of the field-fluctuations
by $g_\DOF$ kinds of the real scalar fields.
According to our mean field approximation of the gravity,
only the interaction with the mean-field gravity is important.
Then the approximation of the scalar fields is justified.
The result similar to ours is expected by 
more proper analysis of graviton, photon, fermion and so on.

The minimal structure model has the entropy $\log g_\DOF$
because the quantum in the ball
belongs to one of $g_\DOF$ kinds of fields. 
If we assume that the Bekenstein's entropy
$S = (1/4) 4\pi r_\BH^2$ in \cite{Bekenstein:1973ur}
is always correct,
the entropy of the ball becomes about $4$ and
the degree of freedom $g_\DOF = \exp S$ becomes about $54$.
While twice of $g_\DOF$ is similar to
the degree of freedom in the Standard Model,
the meaning of this is not clear.

\begin{flushleft}
 {\Large\bf ACKNOWLEDGMENTS}
\end{flushleft}

 I would like to thank
 Ofer~Aharony, Micha~Berkooz,
 Hiroshi~Ezawa, Satoshi Iso,
 Barak~Kol, 
 Masao~Ninomiya and Kunio~Yasue
 for useful discussions.
 I am grateful to Kei~Shigetomi
 for helpful advice and also for careful reading of the manuscript.
 The work has been supported by the Okayama Prefecture.

\section*{Appendix}

As an example
we present
the full expression of a element of the energy-momentum tensor as
\begin{eqnarray}
 T_0^{\ 0}
  &=&
  \sum_{n_1m_1l_1} \sum_{n_2m_2l_2}
  \frac{1}{\sqrt{2 \omega_{n_1}C_{n_1l_1}}}
  \frac{1}{\sqrt{2 \omega_{n_2}C_{n_2l_2}}}\nonumber\\
  && \qquad
   \;\times\;
   \frac{1}{2}
  \Biggl[
   \;+\;  
   \omega_{n_1}\omega_{n_2}
   \frac{1}{F} f_{n_1l_1} f_{n_2l_2}
   Y_{l_1}^{\ m_1} Y_{l_2}^{\ m_2}
   \;+\;
   \frac{1}{G} \partial_r f_{n_1l_1} \partial_r f_{n_2l_2}
   Y_{l_1}^{\ m_1} Y_{l_2}^{\ m_2}\nonumber\\
 && \qquad\qquad\ \ 
   \;+\;
   \frac{1}{r^2} f_{n_1l_1} f_{n_2l_2}
   \left(
    \partial_\theta Y_{l_1}^{\ m_1} \partial_\theta Y_{l_2}^{\ m_2}
    +
    \frac{\partial_\varphi Y_{l_1}^{\ m_1} \partial_\varphi Y_{l_2}^{\ m_2}}
    {\sin^2\theta}    
   \right)
  \Biggr]\nonumber\\
 && \qquad
  \;\times\;
  \left(
     e^{-i(\omega_{n_1} - \omega_{n_2}) t}
     a_{n_1m_1l_1} a_{n_2m_2l_2}^\dagger
   + e^{+i(\omega_{n_1} - \omega_{n_2}) t}
     a_{n_1m_1l_1}^\dagger a_{n_2m_2l_2}
  \right)\nonumber\\
  &+&
  \sum_{n_1m_1l_1} \sum_{n_2m_2l_2}
  \frac{1}{\sqrt{2 \omega_{n_1}C_{n_1l_1}}}
  \frac{1}{\sqrt{2 \omega_{n_2}C_{n_2l_2}}}\nonumber\\
  && \qquad
   \;\times\;
   \frac{1}{2}
  \Biggl[
   \;-\;
   \omega_{n_1}\omega_{n_2}
   \frac{1}{F} f_{n_1l_1} f_{n_2l_2}
   Y_{l_1}^{\ m_1} Y_{l_2}^{\ m_2}
   \;+\;
   \frac{1}{G} \partial_r f_{n_1l_1} \partial_r f_{n_2l_2}
   Y_{l_1}^{\ m_1} Y_{l_2}^{\ m_2}\nonumber\\
 && \qquad\qquad\ \ 
   \;+\;
   \frac{1}{r^2} f_{n_1l_1} f_{n_2l_2}
   \left(
    \partial_\theta Y_{l_1}^{\ m_1} \partial_\theta Y_{l_2}^{\ m_2}
    +
    \frac{\partial_\varphi Y_{l_1}^{\ m_1} \partial_\varphi Y_{l_2}^{\ m_2}}
    {\sin^2\theta}    
   \right)
  \Biggr]\nonumber\\
 && \qquad
  \;\times\;
  \left(
     e^{-i(\omega_{n_1} + \omega_{n_2}) t}
     a_{n_1m_1l_1} a_{n_2m_2l_2}
   + e^{+i(\omega_{n_1} + \omega_{n_2}) t}
     a_{n_1m_1l_1}^\dagger a_{n_2m_2l_2}^\dagger
  \right).
  \label{full-T00.eq}
\end{eqnarray}
Only the first half of the expression \EQ{full-T00.eq}
contributes to the expectation value of the eigenstate
for the Hamiltonian.


\end{document}